\documentstyle[preprint,aps]{revtex}
\tighten
\draft
\begin{document}
\widetext
\preprint{NYU-TH/10/98/06, DFAQ-98/TH/02}
\bigskip
\bigskip
\title{Flavor Violation in Theories with TeV Scale Quantum Gravity}
\medskip
\author{Zurab Berezhiani$^{a}$ and Gia Dvali$^{b}$\footnote{E-mail: 
gd23@is9.nyu.edu}}

\bigskip
\address{$^a$Universit\`a dell'Aquila, I-67010 Coppito, 
L'Aquila, Italy, and \\
Institute of Physics, Georgian Academy of Sciences, Tbilisi, Georgia \\ 
$^b$Physics Department, New York University, 4 Washington Place, 
New York, NY 10003 \\ 
and ICTP, Trieste, Italy}
\date{\today}
\bigskip
\medskip
\maketitle

\begin{abstract}
We study the effects of possible flavor-violating operators 
in theories with the TeV scale quantum gravity, 
in which the ordinary matter is localized on
a $3$-brane embedded in the space with $N$ extra dimensions, 
whereas gravity propagates in the bulk. 
These operators are scaled by the fundamental Planck mass 
$M_{Pf}\sim$ TeV and must be suppressed by the gauge family symmetries. 
We study suppression of the most dangerous and 
model-independent operators. Several points emerge. 
First, we show that the Abelian symmetries can not do the job 
and one has to invoke non-Abelian $U(2)_F$ (or $U(3)_F$) symmetries. 
However, even in this case there emerge severe restrictions 
on the fermion mixing pattern and the whole structure of the
theory.
In order not to be immediately excluded by the well-known bounds, 
the horizontal gauge fields {\it must} be the bulk modes, 
like gravitons. For the generic hierarchical breaking
pattern the four-fermion operators induced by the tree-level 
exchange of the bulk gauge fields are unsuppressed for $N = 2$. 
For $N > 3$ the suppression factor goes as a 
square of the largest $U(2)_F$-non-invariant Yukawa coupling, 
which implies the lower bound $M_{Pf} > 10$ TeV or so 
from the $K^0 - \bar K^0$ system. Situation is different in the scenarios when
flavor Higgs fields (and thus familons) live on a ($3 + N'$)-brane of lower
dimensionality than the gauge fields.
The further suppression of gauge-mediated operators can be achieved by an explicit
construction: for instance, if $U(2)_F$ is
broken by a vacuum expectation value of the doublet, 
the troublesome operators are suppressed 
in the leading order, due to custodial $SO(4)$ symmetry 
of the Higgs-gauge quartic coupling.
\end{abstract}
\narrowtext

\subsection{Introduction}

 One of the fundamental mysteries of the Nature 
is the enormous hierarchy between the observable values of the 
weak interaction scale $M_W$ and the Planck scale $M_P$.
A possible solution to this mystery \cite{add} may have to do 
with the fact that the fundamental scale of gravitational interaction 
$M_{Pf}$ is as low as TeV, whereas the observed weakness 
of the Newtonian coupling constant $G_N \sim M_P^{-2}$ is 
due to the existence of $N$ large ($ \gg {\rm TeV}^{-1}$) 
extra dimensions into which the gravitational flux can spread out. 
At the distances larger than the typical size of these extra dimensions
($R$) gravity goes to its standard Einstein form. 
For instance, for two test masses separated by the distance 
$r\gg R$, the usual $1/r^2$ Newtonian low is recovered, and  
the relation between the fundamental and observed Planck scales 
is given by:
\begin{equation}
M_P^2 = M_{Pf}^{N + 2}R^N
\end{equation}
In such a theory, quantum gravity becomes strong at energies $M_{Pf}$, 
where presumably all the interactions must unify.\footnote{Witten suggested
\cite{witten} that string scale may be around the scale of
supersymmetric unification $\sim 10^{16} GeV$. The possibility of
having an extremely low string scale was also discussed by Lykken
\cite{lykken}.} 
For all the reasonable choices of $N$, the size of extra radii is
within the experimental range in which the strong and 
electroweak interactions have been probed. 
Thus, unlike gravity the other observed particles should not
"see" the extra dimension (at least up to energies $\sim$ TeV). 
In ref. \cite{add} this was accomplished by postulating 
that all the standard model particles are confined to
a $3 + 1$-dimensional hyper-surface ($3$-brane), whereas gravity 
(as it should) penetrates the extra dimensional bulk\cite{walluniverse1}-\cite{kt}. 
Thus, on a very general grounds, the particle spectrum of
the theory is divided in two categories: 
(1) the standard model particles living on the $3$-brane
(brane modes); 
(2) gravity and other possible hypothetical particles 
propagating in the bulk (bulk modes).
Since extra dimensions are compact, any $4 + N$-dimensional 
bulk field represents an infinite tower of the 
four-dimensional Kaluza-Klein (KK) states with masses quantized in
units of inverse radii $R^{-1}$. 
An important fact is that each of these states 
(viewed as a four dimensional mode) has extremely weak, 
suppressed at least as $M_P^{-1}$ couplings to the brane modes. 
The ordinary four-dimensional graviton, which is
nothing but a lowest KK mode of the bulk graviton,  is a
simplest example. In what follows, this fact will play a 
crucial role as far as the other possible bulk particles 
are concerned. 

Obviously, such a scenario requires various compatibility 
checks many of which were performed in
\cite{add},\cite{aadd},\cite{add*},\cite{adm}\cite{ndm}\cite{giudice},\cite{ns}.
It was shown that this scenario passes a 
variety of the laboratory and astrophysical tests. 
Most of the analysis was mainly concerned 
to check the "calculable" consequences of the theory, 
ones that obviously arise and are possible
to estimate in the field (or string) theory picture. 
On the other hand, there are constraints based on the 
effects whose existence is impossible to proof or rule out
at the given stage of understanding the quantum gravity, 
but which are usually believed to be there. 
An expected violation of global quantum numbers by gravity 
is an example. There are no rigorous proofs of such effect, 
nor any knowledge of what their actual strength should be. 
Yet, if an effect is there with a most naive dimensional
analysis we expect it to manifest itself in terms of all possible 
gauge-symmetric operators suppressed by the Planck scale.  
Below we adopt this philosophy, which then imposes 
the severe constraints on the proposal of ref. \cite{add}, 
since now the fundamental gravity scale $M_{Pf}$ 
is as low as TeV! 
Issues regarding baryon and lepton number violation
were discussed in \cite{aadd},\cite{tye}, \cite{add*} and 
some ways out were suggested\footnote{The unification of gauge couplings is another
issue \cite{diduge} not to be addressed in this paper.}.
In the present paper we will discuss the flavor problem 
in TeV scale quantum gravity theories induced by   
higher order effective operators cutoff by the scale 
$M_{Pf}$ 
that can contribute to various flavor-changing neutral 
processes (FCNP), like $\bar K^0-K^0$ or $D^0 - \bar D^0$ 
transitions, $\mu \rightarrow e\gamma$ decay etc.

It is normal that the flavor-violating interactions provide 
severe constraints to any new physics beyond the 
standard model.\footnote{ 
Unlike the baryon number non-conservation, 
gravity-mediated FCNP are only important for a very low scale 
quantum gravity theories: the lowest dimensional 
baryon-number-violating operators scaled as $M_{Pf}^{-1}$ 
are problematic even for theories with $M_{Pf} = M_P$, 
whereas the flavor problem disappears already for 
$M_{Pf} > 10^{(7-8)}$ GeV or so.}  
As we will see below the flavor problem provides 
severe constraints both on the symmetry structure of 
the theory and on the structure of fermion mass  matrixes.

\subsection{The Problematic Operators and Gauge Family Symmetries}

As said above, in the effective low energy theory below TeV, we expect all possible
flavor
violating four-fermion operators scaled by $M_{Pf}^{-2}$. Some of these give
unacceptably large contributions to the flavor changing processes and must be
adequately suppressed. Let us consider what are the symmetries that can do the
job\footnote{
$N=1$ supersymmetry can not be of much help due to the following reasons:
first in any case it must be broken around TeV scale, and secondly the four-fermion
interaction can arise from the K\"ahler metric, which can not be controlled by
holomorphy.}. Usually one of the most sensitive processes to a new flavor-violating
physics
is the $K^0 - \bar K^0$ transition. Corresponding effective operator in the present
context
would have a form
\begin{equation}
 {(\bar s d)^2 \over M_{Pf}^{2}} \label{kaon}
\end{equation}
This can only be suppressed by the symmetry that acts differently on $s$ and $d$ and
therefore is a
{\it family} symmetry. Thus, as a first requirement we have to invoke a gauge family
symmetry.
In an ordinary (four-dimensional) field theory there would be an immediate problem with
this proposal. In order to adequately suppress the operator (\ref{kaon}), the
symmetry in
question
should be broken (well) below TeV. But there are well known lower bounds\cite{bounds}
$>> TeV$ on the scale of gauge flavor symmetry breaking. This bound comes
from a tree-level exchange of the horizontal gauge boson, that
will mediate the same FCNP for which the symmetry was invoked!
However, one has to remember that this is true as far as the four-dimensional field theory
is
concerned. Recall that in our case there are two type of particles, ordinary particles
living on a brane and the bulk modes. If the horizontal gauge field is the bulk mode
the situation is different. Now the coupling of each KK excitation
to the ordinary particles will be enormously suppressed.
This saves the scenario from being {\it a priory} excluded.
The large multiplicity of the
exchanged KK states however works against us and at the end puts severe
constraint on the dimensionality of extra space, flavor breaking scale and the pattern
of quark masses. We will discuss this in detail below.

 Now let us discuss what are the symmetries that one can use.
In the limit of zero Yukawa couplings the standard model exhibits 
an unbroken flavor symmetry group
\begin{equation}
G_F = U(3)_{Q_L}\otimes U(3)_{u_R}\otimes U(3)_{d_R}\otimes U(3)_{l_L}\otimes U(3)_{e_R}
\end{equation} 
If one is going to gauge some subgroup of $G_F$, 
the Yukawa coupling constants are to be understood as the 
vacuum expectation values of the fields that break this
symmetry \cite{flavor}. 
That is the fermion masses must be generated by the higher dimensional
operators of the form:
\begin{equation}
 \left ({\chi \over M}\right )^N_{ab} H \bar Q_L^aq_R^b \label{chi}
\end{equation}
where $\chi$ are flavor-breaking Higgses. 
To take advantage of the problem, it is natural
and most economical to assume that the above desired operators 
are generated by the same physics which induces the problematic ones. 
Thus we adopt that $M \sim M_{Pf}$.
 An observed fermion mass hierarchy then 
is accounted by hierarchical breaking of $G_F$. 
In the present paper we will not be interested how 
precisely such a hierarchy of VEVs is generated, 
but rather will look for its consequences as far as FCNP are
concerned. 

Now the large Yukawa coupling of the
top quark indicates that at least 
$U(3)_{Q_L}\otimes U(3)_{u_R} \rightarrow
U(2)_{Q_L}\otimes U(2)_{u_R}$ breaking should occur 
at the scale $\sim M_{Pf}$ and thus it can not provide any 
significant suppression. 
Therefore, the selection rules for the operators
that involve purely $Q_L$ and $u_R$ states can be based essentially 
on $U(2)_{Q_L}\otimes U(2)_{u_R}$ symmetry or its subgroups. 
The most problematic dimension six operators in this respect is 
(below we will not specify explicitly a Lorenz structure, 
since in each case it will be clear from the context): 
\begin{equation}
 (\bar Q_L^aQ_{La})(\bar Q_L^{b}Q_{Lb})~+~ 
(\bar Q_L^aQ_{Lc})(\bar Q_L^b Q_{Ld})
\epsilon_{ab}\epsilon^{cd}  \label{problematic} 
\end{equation}
They both give a crudely similar effect. 
So let us for definiteness concentrate on
the second one. Written in terms
of initial $s$ and $d$ states (call it 'flavor basis') it has a form:
\begin{equation}
 (\bar s_Ls_L)(\bar d_Ld_L) - (\bar s_Ld_L)(\bar d_Ls_L)
\end{equation}
In general initial $s$ and $d$ states are not physical states
and are related to them by $2\times 2$ rotation $D_L$, which diagonalizes $1-2$ block of
the
down quark mass matrix  $M^d$. The problem is that $D_L$ is {\it not} in general unitary
due
to non-zero $1-3$ and/or $2-3$ mixing in $M^d$. Note that this
elements can be of the order of one, without conflicting with small
$2-3$ and $1-3$ mixings in the Cabibbo-Kobayashi-Maskawa (CKM) matrix, since CKM
measures a mismatch between rotations of $u_L$ and $d_L$ and not
each of them separately.
If so,  then in the physical basis the disastrous operator
\begin{equation}
 (\bar s_Ld_L)(\bar s_Ld_L)     \label{induced}
\end{equation}
will be induced with an unacceptable strength. This puts a severe constraint on the
structure of $M^d$. In particular all the anzatses with both large $1-2$ and  $2-3$ (or
$1-3$) elements are ruled out. Note that smallness of $1-3$ and $2-3$ mixing in
$M^d$ also works in favor of suppression of $B^0 - \bar B^0$ transitions from the same
operator. Analogously, unitarity of $1-2$ diagonalization in $M^u$ suppresses the
$D^0 - \bar D^0$ transition. In this respect the safest scenario would be the one in which,
$1-2$ mixing in CKM comes mostly from down type masses, whereas $2-3$ mixing from ups.

Much in the same way the operator
\begin{equation}
 (\bar d_R^ad_{R\alpha})(\bar d_R^bd_{R\beta})
\epsilon_{ab}\epsilon^{\alpha\beta}   \label{RRRR}
\end{equation}
gives unitarity constraint on a $1-2$ block of $D_R$, 
which can be somewhat milder since the operator (\ref{RRRR}) 
can in principle be suppressed by $U(3)_{d_R}$-symmetry,
by a factor $\sim m_b/m_t$.

For the operators which involve both left and right-handed quark states, the
suppression factors are more sensitive 
to what subgroup of the $G_F$ is gauged.

\subsection{$L\times R$-Type Symmetries}

If the left and the right-handed quark are transforming 
under different $U(2)_{FL}\otimes U(2)_{FR}$
flavor symmetries. The only possible unsuppressed operator is
\begin{equation}
 (\bar Q_L^aQ_{La})(\bar Q_R^bQ_{R_b}) \label{barLR}
\end{equation} 
(plus its Fierz-equivalent combinations).
Again, this is harmless only if $D_L$ and $D_R$ are nearly unitary, 
which brings us back to the constraint discussed above.

\subsubsection{The Diagonal $U(2)_F$} 

From the point of view of an anomaly cancellation,
the most economic possibility would be to gauge a diagonal 
subgroup of $G_F$ under which all fermions are in the 
fundamental representation. 
Then at scales below $M_{Pf}$ we are left with an effective 
$U(2)_F$ symmetry. In such a case one encounters an option 
whether $U(2)_F$ is a chiral or vector-like symmetry. 
It turns out that the chiral $U(2)_F$  is inefficient to
suppress a large flavor violation, whereas the vector-like 
one can do the job, provided a stronger
restriction on the fermion mass pattern is met.\footnote{
Unfortunately, however, the vector-like 
flavor symmetry $U(2)_F$ allows the fermion mass 
degeneracy, and unnatural 
conspiracies would be needed for explaining their observed 
splitting. In this view, it would be most natural if the 
vector-like $U(2)_F$ is supplemented by some (discrete of continuous) gauge chiral
piece of the full chiral $(L\times R)$ symmetry.}  

To support the first statement it is enough to consider an operator:
\begin{equation}
 (\bar Q_L^ad_L^{\alpha})(\bar Q_{Ra}d_{R\alpha}) \label{chiralLR}
\end{equation}
This is invariant under the chiral-$U(2)_F$ times an 
arbitrary combination of extra $U(1)$-factors. 
Obviously this operator is a disaster, since it directly contains
an unsuppressed four-Fermi interaction (\ref{kaon}). 
Thus chiral $U(2)_F$ cannot protect us. 
On the other hand the vector-like $U(2)$ 
suppresses (\ref{chiralLR}). Analogous
(self-conjugate) operators in this case would have the form:
\begin{equation}
 (\bar Q_L^ad_{Ra})(\bar d_R^bQ_{Lb}), ~~~~
 (\bar Q_L^ad_{R\alpha})(\bar d_R^bQ_{L\beta})
\epsilon_{ab}\epsilon^{\alpha\beta}   \label{vectLR}
\end{equation}
which contains no (\ref{kaon}) term in the flavor basis. 
The requirement that it will not be induced in the physical 
basis simply translates as a requirement that
\begin{equation}
D_LD_R^+ = D_LD_L^+ = D_R D_R^+ = 1 \label{unitarity}
\end{equation}
 with a great accuracy.
In other words, this means that the fermion mass matrices 
should be nearly Hermitian.\footnote{This might be rather 
natural in the context of the left-right symmetric model 
$SU(2)_L\times SU(2)_R\times U(1)$}
If this is satisfied other operators do not cause
an additional constraints, since they are further suppressed.
For instance, non-self-conjugate operators:
\begin{equation}
(\bar Q_L^ad_{Ra})(\bar Q_L^bd_{Rb}), ~~~~
(\bar Q_L^ad_{R\alpha})(\bar Q_L^bd_{R\beta})
\epsilon_{ab}\epsilon^{\alpha\beta}   \label{vectLR}
\end{equation}
carry two units of weak isospin and must be 
suppressed by extra factor $\sim \left ( {M_W \over M_{Pf}} \right )^2$.
In conclusion, we see that all working versions converge to the 
requirement (\ref{unitarity}).

\subsubsection{Why Abelian Symmetries Cannot work?}

Although our analysis was quite general, 
one may wonder whether by considering non-Abelian symmetries, 
one is not restricting possible set of solutions: for example
requirement of $SU(2)$ symmetry restricts the possible charge 
assignment under the additional $U(1)$ factors which otherwise 
could be used for the same purpose. In other words,
one may ask, whether instead of non-Abelian symmetries one 
could have invoked a variety of $U(1)$ factors and by properly 
adjusting charges of the different fermions get the same
(or even stronger) suppression of FCNP. 
We will argue now that this is not the case,
and even (neglecting esthetics and various technical 
complications, like anomalies) if one allows completely arbitrary 
charge assignment under an arbitrary number of $U(1)$-factors,
the problem cannot be solved. The reason for this is that 
no Abelian symmetry can forbid the operators of the form:
\begin{equation}
 C_{ab}(\bar Q_L^aQ_{La})(\bar Q_L^bQ_{Lb})
\end{equation}
and similarly for right-handed fermions.
Since no non-Abelian symmetry is invoked the coefficients 
$C_{ab}$ are completely arbitrary. Due to this fact there is 
no choice of fermion mass matrixes which would avoid appearance 
of either $(\bar sd)^2$ or $(\bar uc)^2$ unsuppressed vertexes. 
Since $C_a$ are arbitrary, non-appearance of any of these operators 
would mean that the flavor and the physical quark states are equal 
(that is the masses are diagonal in a flavor basis). But this is
impossible to be the case in both up and down sectors 
simultaneously due to non-zero Cabibbo mixing $\sin\theta_C=0.22$.  
Thus at least one of these operators should be induced and the
suppression factor cannot be smaller than
\begin{equation}
 \sin\theta_C^2/M_{Pf}^2
\end{equation}
This gives rise to an unacceptably large contribution  
to either $\bar K^0 - K^0$ or $\bar D^0 - D^0$ transitions.
The only way to avoid the problem would be a conspiracy
between the $C_{ab}$ coefficients, which can be guaranteed 
by non-Abelian $U(2)$ symmetry,
(subject to unitarity of $U_L, U_R, D_L$ and $D_R$ transformations).

\subsection{Electroweak Higgs-Mediated Flavour Violation.}

Existence of the $M_{Pf}$-suppressed operators 
brings another potential source of flavour violation, 
mediated by the electrically neutral component ($H^0$) 
of the standard model Higgs.
In the standard model this sourse is absent since 
the couplings of $H^0$ are automatically
diagonal in the physical basis. 
This is not any more true if higher dimensional operators
with more Higgs verteces are involved\cite{Rattazzi}. 
For instance, add the lowest possible such operator
\begin{equation}
 \left (g_{ab} +  h_{ab}{H^+H \over M_{PF}^2 +...} \right ) H\bar Q^a_Lq_{Rb}
\end{equation}
where $g_{ab}$ and $h_{ab}$ are constants. 
After $H$ gets an expectation value the fermion masses become
\begin{equation}
M_{ab} =  \left (g_{ab} +  h_{ab}{|\langle H \rangle|^2 
\over M_{PF}^2} + ...\right ) \langle 
H \rangle
\end{equation}
whereas the Yukawa couplings of the physical Higgs are
\begin{equation}
Y_{ab} = g_{ab} +  3h_{ab}{|\langle H \rangle|^2 \over M_{PF}^2 +... } 
\end{equation}
In the absence of flavour symmetries the matrixes $g_{ab}$ and $h_{ab}$ are arbitrary $3\times
3$ matrices and thus $M_{ab}$ and $Y_{ab}$ are not diagonal in the same basis. This induces
an unacceptably large flavour violation. In the present context however according to
Eq(\ref{chi}), $g_{ab}$ and $h_{ab}$ must be understood as
the VEVs of the horizontal Higgs scalars
\begin{equation}
g_{ab} \sim h_{ab} \sim \left ({\chi \over M}\right )^N_{ab} \label{hierarchy}
\end{equation}
and thus obey an {\it approximately same} hierarchy \cite{gd}. This can reduce the
resulting flavor violation to an acceptable level. For instance, adopting
anzats $Y_{ab} \sim {\sqrt{m_am_b} \over M_W}$\cite{chengsher}, where $m_a$ are masses of
physical
fermions, the resulting flavour violation can be below the experimental limits.

\subsection{decay $\mu \rightarrow e\gamma$, etc.}

The suppression of lepton-flavour violating processes through dimension-$6$ operators
goes much in the same spirit as discussed above for quarks. The constrains on
the $M^l$ mixing angles can be satisfied easier since not much is known about the
lepton mixing angles.
 Dimension five operators can be more problematic. 
For instance the lowest operator
inducing $\mu \rightarrow e\gamma$ transition is:
\begin{equation}
{H \over M_{PF}^2} \bar e \sigma^{\mu\nu}\mu F_{\mu\nu} 
\end{equation}
This has the same chirality structure as the $m_{\mu e}$ mass term, and thus we expect to be
suppressed
by the same flavour symmetry that guarantees its smallness. 
An exact strength of the suppression factor is very sensitive to the mixing in
the charged lepton matrix and can be as large as $m_{\tau} \over M_{Pf}^2$ for
the maximal $e-\mu -\tau$ mixing angles. But can be zero if mixing is absent.

The experimental limit 
${\rm Br}(\mu \rightarrow e\gamma) < 5\times 10^{-11}$ translates 
into 
\begin{equation}
\lambda_{e\mu} < 3\times 10^{-6} \cdot
\left({M_{Pf}\over 1~{\rm TeV}} \right)^2  
\end{equation} 
which is satisfied for $\lambda_{e\mu}\sim 
\sqrt{m_em_\mu}/\langle H\rangle = 4\cdot 10^{-5}$ and 
$M_{Pf}\sim 3$ TeV. Analogously, for the $\tau \rightarrow \mu\gamma$ 
decay ${\rm Br}(\tau \rightarrow \mu\gamma) < 3\times 10^{-6}$ 
translates into 
\begin{equation}
\lambda_{\mu\tau} < 3\times 10^{-2} \cdot
\left({M_{Pf}\over 1~{\rm TeV}} \right)^2  
\end{equation} 
which is well above the geometrical estimate 
$\lambda_{\mu\tau}=
\sqrt{m_\mu m_\tau}/\langle H\rangle = 2.5\cdot 10^{-3}$.  

Somewhat stronger constraints come from the electron and 
neutron EDMs. For example, the experimental limit 
$d_e < 0.3\cdot 10^{-26}~ e\cdot$cm implies 
\begin{equation}
{\rm Im}\lambda_{e} < 10^{-9} \cdot
\left({M_{Pf}\over 1~{\rm TeV}} \right)^2  
\end{equation} 
therefore, for $\lambda_e$ taken of the order of the electron 
Yukawa coupling constant, $\lambda_e \sim m_e/\langle H\rangle \sim 
10^{-6}$ with a phase order 1, one needs to take 
$M_{Pf} > 30$ TeV or so. Analogous constraint emerges 
from the light quarks (i.e. neutron) EDM.

\subsection{Gauging Flavor Symmetry in the Bulk}.
Up to now we were discussing suppression of $M_{Pf}$-cutoff 
operators by the gauge flavor symmetries. 
What about flavor-violation mediated by the horizontal gauge
fields? Naively, one encounters a puzzle here: in order to suppress
quantum-gravity-induced operators, $G_F$ must survive 
at scales below $M_{Pf}$, but in this case
the gauge bosons can themselves induce problematic operators. 
Situation is very different if the horizontal gauge 
bosons are the bulk fields.

 Generic procedure of gauging an arbitrary gauge symmetry 
in the bulk was discussed in details in \cite{add*} and 
it was shown that: 1) an effective coupling of the
bulk gauge-field and its KK partners to the brane modes 
is automatically suppressed by $g_4 \sim M_{Pf}/M_P$ and, 
2) whenever the symmetry is broken on the brane
(only)
the mass of the gauge field is suppressed by $\sim M_{Pf}/M_P$ 
independently of the number of extra dimensions. 
In the other words, as it should be, the symmetry broken on
the brane is "felt" by the brane fields much 
stronger then by the bulk modes. This is not
surprising, since the bulk modes "spent" much more 
time in the bulk where symmetry is
unbroken.

 Consider a gauge field of some symmetry group 
$G$ propagating in the bulk. We will assume the scale
of the original
$4 + N$ - dimensional gauge coupling to be 
$g_{(4 + N)} \sim M_{Pf}^{-{N \over 2}}$.
From $4$-dimensional point of view, this 
gauge field represents an infinite number of KK
states out of which only the zero mode $A_{\mu}^0$ 
shifts under the $4$-dimensional local
gauge transformation, whereas its KK partners are massive states.
 All these states couple to the gauge-charged matter localized on
the brane through an effective
four-dimensional gauge coupling
\begin{equation}
g_4^2 \sim 1/(RM_{Pf})^N \sim M^2_{Pf}/M_P^2
\end{equation}
Consequently if any of the Higgs scalars localized 
on the brane gets nonzero VEV
$\langle \chi (x_A)\rangle = \delta(x_A - x_A^0)\langle \chi \rangle$ 
(where $x_A^0$ are coordinates of the brane) 
all the bulk states get a minuscule mass shift
\begin{equation}
\delta m^2 \sim \langle \chi \rangle^2 M^2_{Pf}/M_p^2 \label{wallshift}
\end{equation}
On the other hand if the Higgs scalar is a 
bulk mode and its VEV is not localized on the
brane, but rather is constant in the bulk, 
the gauge fields get an unsuppressed mass shift
\begin{equation}
\delta m^2 \sim \langle \chi \rangle^2/ M^N_{Pf} \label{bulkshift}
\end{equation}
Note that the bulk scalar $\chi$ has 
dimensionality of $({\rm mass})^{1 + N/2}$.

What is the implication of these facts for the flavor symmetry? 
Let $G_F = U(3)_F$ be a vector-like family symmetry under 
which all the fermions are triplets and let us estimate
flavor violation induced by exchange of its gauge field. 
Consider for instance $M^0 - \bar M^0$ transitions 
(where $M = K, B, D$). Obviously, if $U(3)_F$ was unbroken,
all the four-fermion operators induced by their exchange 
would never contribute to any of
these processes, since the only possible invariant is
\begin{equation}
    \bar q^a q_a \bar q^b q_b  \label{flavor-conserving}
\end{equation}
However, $U(3)_F$ must be broken in order to account for 
the hierarchy of fermion
masses. Let $\chi$ be the Higgs that does this breaking. 
We can consider three options:

 {\bf 1. Breaking occurs only on the brane}. 
Unfortunately this option is ruled out due to the following reason. 
If $U(3)_F$ is broken on the brane, then there must be
massless pseudo-scalar modes localized on it. 
These are Goldstone bosons (familons \cite{familon}) of broken
$U(3)_F$ which have both flavor-diagonal and 
flavor-non-diagonal couplings to the
ordinary matter, suppressed by $\langle \chi \rangle$. 
This is excluded due to various astrophysical and 
laboratory reasons \cite{familon,familonruled}.
At a first glance, in the present context 
this statement may appear as a surprise,
since by assumption $U(3)_F$ is a gauge symmetry and 
thus troublesome familons must be eaten
up by the gauge fields. Recall however that the gauge coupling is
abnormally small (and the scale of symmetry breaking is not large). 
In such a situation it is more useful to argue in terms of 
the massless Goldstones, rather than massive gauge bosons.\footnote{
The situation is analogous to the case of the low energy 
supersymmetry, when the dominant coupling of gravitino is provided 
by its goldstino component!} 
Thus we are left with the following option.

 {\bf 2. Breaking occurs in the bulk}. 
In this case familons are the bulk modes and are
totally safe by the same reason as the bulk gravitons \cite{add*}. 
Thus the dominant contribution to the flavor-violation is provided 
by the gauge components. Let us estimate this
contribution to the effective four-Fermi operators 
mediating $M^0 - \bar M^0 $ processes. This comes from
the tree-level exchange of infinite tower of $KK$ states. 
The mass of each individual KK mode is 
\begin{equation}
m^2_K = \langle \chi \rangle^2/ M^N_{Pf} + {|n|^2 \over R^2}
\end{equation}
where $|n| = \sqrt{n_A^2}$ and $n_A$ are integers. 
The second contribution is flavor-universal. 
Thus for the heavy states the flavor violation will be suppressed by
$\sim \langle \chi \rangle^2 R^2/ M^N_{Pf}|n|^2$. 
Note that the same scalars are responsible for the fermion masses through
\begin{equation} 
 \int dx^{4 + N} \delta(x_A){\chi_{b}^a \over 
M_{Pf}^{1 + {N \over 2}}} H \bar Q_{La}q_R^b     \label{chibulk}
\end{equation}
where dimensionality of the denominator comes from the fact 
that $\chi$ is a bulk mode. Thus
$U(3)_F$-violating mass can be parameterized as
\begin{equation}
\langle \chi \rangle^2/ M^N_{Pf} = (\lambda M_{Pf})^2
\end{equation}
where $\lambda$ is roughly the Yukawa coupling of the fermion 
(e.g. for $U(2)_H$ gauge bosons $\lambda$ can be taken to be 
$\sim m_c/\langle H\rangle \sim 10^{-2}$ or so).
It is useful to evaluate contributions of the modes 
${|n|^2 \over R^2} < (\lambda M_{Pf})^2$ and 
${|n|^2 \over R^2} > (\lambda M_{Pf})^2$ separately.
Each of the first states generates an operator scaled by 
${g_4^2 \over (\lambda M_{Pf})^2}$
whereas their multiplicity is roughly 
$\sim (\lambda M_{Pf}R)^N$. Therefore their
combined effect gives an effective four-fermion regulator
\begin{equation}
\sim {\lambda^{N - 2} \over M_{Pf}^2}. \label{scaling}
\end{equation}
The flavor violating contribution of the modes with 
${|n|^2 \over R^2} > (\lambda M_{Pf})^2$ is crudely 
given by the sum
\begin{equation}
\sum_{n_A} {g_4^2R^4 \over |n|^4} (\lambda M_{Pf})^2 
\sim g_4^2R^4 (\lambda M_{Pf})^2 |n|^{N - 4} |_{min}^{max}
\end{equation}
and it is power-divergent for $N > 4$.
Cutting off from above this sum at 
$|n|_{{\rm max}} \sim (M_{Pf}R)$ we get that the
amplitude goes as
\begin{equation}
\sim {\lambda^2 \over M_{Pf}^2} \left ( 1 - \lambda^{N - 4} \right ) \label{scaling2}
\end{equation}
Thus we see that the dominant contribution comes from the lowest modes.
Combining everything we get that for $N > 3$ the operators scale as
\begin{equation}
\sim {\lambda^2 \over M_{Pf}^2} 
\end{equation}
and are problematic even if 
$\lambda \sim 10^{-2}-10^{-3}$.
The case $N < 4$ is even more problematic and, 
in particular, there is no suppression for $N = 2$.
At a first glance this may appear as a surprise: 
since in the limit $\lambda \rightarrow 0$ the transition 
should be absent. Note however, that this does not
contradict to above result, since for 
$\lambda M_{Pf}$ becoming smaller than the meson
mass $m_{M}$ an additional power suppression 
$\sim {\lambda M_{Pf} \over m_{M}}$ must appear 
in the flavor violating transition amplitude. 
 Thus, we are lead to the third option.

{\bf 3. Gauge fields and familons on the branes of different dimensionality.}
Imagine that flavor Higgs fields are not $3$-brane modes and live in space of larger
dimensionality. But,
unlike gauge fields,
they can only propagate in the bulk of less $N'< N$ dimensions. That is assume that
they live on $3 + N'$-brane which contains our $3$-brane Universe as a sub-space.
When the breaking occurs on the $3 + N'$-brane, both, masses of bulk
gauge fields and couplings of familons to the standard model fermions will
be suppressed by volume factors. The question is whether for certain
values of $N'< N$, both gauge and familons-mediated processes can be adequately
suppressed. Consider first some gauge-mediated FCNP. Since $\chi$-s are $3 +
N'$-dimensional
fields, according to (\ref{chibulk}) their VEV can be parameterized as
\begin{equation}
\langle \chi \rangle^2/ M^{N'}_{Pf} = (\lambda M_{Pf})^2
\end{equation}
The resulting flavor-non-universal mass for each gauge KK mode is
\begin{equation}
m_{fv}^2 = {(\lambda M_{Pf})^2 \over (M_{Pf}R)^{N - N'}}
\end{equation}
or if translated in terms of $M_P$, $m_{fv}^2 = 
(\lambda M_{Pf})^2 (M_{Pf}/M_P)^{{2(N - N') \over N}}$. For the gauge-mediated
flavor-violation to be suppressed, this should be smaller than the typical
momentum transfer in the process $M^0 - \bar M^0$. If this is the case,
then the contribution to the process from the light ($<< m_M$) and heavy
($>> m_M$) modes go as $m_{fv}^2m_M^{N - 4}/M_{Pf}^N$ and 
$m_{fv}^2/M_{Pf}^4$ respectively and are suppressed. Now let us turn to the familon
couplings. Since $\chi$ are $4 + N'$ dimensional fields, so are the familons and their
effective decay constant is\cite{add*}
\begin{equation}
1/(\lambda M_{Pf})(M_{Pf}R)^{N'/2}
\end{equation}
Again because of the bulk-multiplicity factor their emission rate is
amplified. For instance the star-cooling rate becomes\cite{add*}
\begin{equation}
\sim (\lambda M_{Pf})^{-2}(T/M_{Pf})^{N'}
\end{equation}
where $T$ is the temperature in the star. This is safe for $N' > 2$ even for $M_{Pf}
\sim$TeV. Contribution from light and heavy modes to familon-mediated flavor-violating
amplitudes are suppressed as
\begin{equation}
{1 \over (\lambda M_{Pf})^2}(m_M /M_{Pf})^{N'}
\end{equation}
and
\begin{equation}
{\lambda^{N'- 4}m_M^2 \over  M_{Pf}^4}
\end{equation}

Finally let us consider flavor-violating operators induced by
horizontal Higgses. This can be analyzed much in the same way
as was done above for the gauge fields. If the non-zero VEV occupies the
whole $3 + N'$-brane volume  where Higgs can freely propagate, then the
resulting dangerous operators are scaled as the largest of (\ref{scaling}) and
(\ref{scaling2})
where now $N$ must be understood as the number of the dimensions
where Higgs can propagate. This is very much like gauge-contribution
in the case of bulk-breaking. The difference is that horizontal Higgses have an extra
suppression factor $\sim (m_W/M_{Pf})^2$ and therefore are relatively safer. 

\subsubsection{Custodial $SO(4)_F$} 

 We must stress that there may very well be the group-theoretical 
cancellations which can weaken the above constraints.
For instance, imagine that $U(2)_F$ symmetry is broken by a 
doublet VEV  $\chi_a$. Gauge boson masses generated in this 
way are automatically $SU(2)$ invariant due to
the {\it custodial} 
global $SO(4)$ symmetry (just like in the standard model)
of the coupling
\begin{equation}
   g^2 (\chi^{*a}\chi_a)A_{\nu}^\alpha A^{\nu\alpha}
\end{equation}
As a result in the leading order $U(2)_F$ 
non-invariant operator structure must cancel out.

\subsection{Conclusions}

Adopting philosophy that the quantum gravity explicitly breaks global symmetries
via all possible operators scaled by powers of $M_{Pf}^{-1}$, we studied some
implications of this
fact for the flavor-violation in theories with TeV scale quantum gravity\cite{add}.
In these theories the ordinary fermions are localized on a $3$-brane embedded in
space with $N$ new dimensions. We have discussed most dangerous
and model independent operators and their suppression by gauged family symmetries.
Non-Abelian symmetries (such as $U(2)_F$) broken below TeV seem to be necessity in this
picture, but in no way they are sufficient for FCNP-suppression. Additional constraints
come out for the structure of the fermion mass matrices and the high-dimensional bulk
properties of the
horizontal gauge fields. All the "safe" versions seem to converge to the structures in
which,
at best, only two generations can have significant mixing per each mass matrix. In
particular, this rules out all possible "democratic" structures: when mixing is maximal
among all three families. 

 To suppress gauge-mediated flavor violation and avoid standard bounds on the scale of
flavor symmetry breaking, the horizontal symmetry should be gauged in the bulk.
If breaking occurs in the bulk, the flavor violation is somewhat reduced
only for large enough number of new dimensions ($N > 2$). On the other hand if breaking
occurs in a subspace with $N'< N$ the gauge-mediated contribution can be strongly
suppressed, but unless $N'$ is also large, the would be familons, that are localized
on a $3 + N'$-brane, can mediate unacceptable flavor-violation.
 
 Combining all the potential sources, it seems that unless implementing an extra source
of suppression "by construction" FCNP are pretty close to their experimental limits.

\acknowledgments

We thank Denis Comelli, Glennys Farrar and  Gregory Gabadadze for very
useful discussions. We learned from Savas Dimopoulos about the complementary
paper\cite{savasnima}, we thank him for discussions. Before sending this paper we
became aware
of number of new papers\cite{new1}-\cite{new4} on phenomenology and cosmology of TeV
scale quantum gravity.

\end{document}